\documentclass[conference,compsoc]{IEEEtran}

\usepackage{graphicx}
\usepackage{amsmath}
\usepackage{amssymb}
\usepackage{cite}
\usepackage{booktabs}
\usepackage{multirow}
\usepackage{url}
\usepackage{comment}
\usepackage{tabularx}
\usepackage{balance}

\title{An Evidence-First Multi-LLM Framework for Auditable Critical-Infrastructure Dependency Modeling}

\author{
\IEEEauthorblockN{Nurjahan, Mst Eshita Khatun, Lamine Noureddine, and Aisha Ali-Gombe}
\IEEEauthorblockA{
Division of Computer Science and Engineering\\
Louisiana State University\\
Baton Rouge, LA, USA\\
\{nurja1, mkhatu3, lnoureddine, aaligombe\}@lsu.edu
}
}

\begin{document}

\maketitle

\begin{abstract}
Critical-infrastructure knowledge is distributed across heterogeneous, incomplete, and weakly structured evidence, making dependency models difficult to construct automatically and difficult to trust. Large language models (LLMs) can extract structured knowledge from such evidence, but direct LLM-to-graph generation risks unsupported relationships, inconsistent terminology, incorrect entity identities, and erroneous dependency endpoints. We present an evidence-first multi-LLM framework for constructing Infrastructure Knowledge Bases (IKBs) and Infrastructure Dependency Graphs (IDGs) from heterogeneous infrastructure documentation. Multiple open-weight LLMs independently extract candidate entities and dependencies from normalized evidence, after which the framework separates evidence verification, ontology grounding, entity resolution, dependency alignment, validation, fusion, and human review. Evidence support, ontology reconciliation, endpoint resolution, model agreement, and human validation remain distinct states, while provenance and unresolved cases are preserved throughout. The validated IKB is then projected deterministically into the IDG without introducing new LLM-generated knowledge. Evaluation in nine infrastructure projects shows that entity recovery achieves substantially higher recall than complete directed dependency recovery and that canonical endpoint resolution is a major constraint in dependency construction. Cross-model overlap is also much lower for dependencies than for entities, indicating that the models often produce non-overlapping candidate assertions rather than a stable majority consensus. These findings support an auditable evidence-to-IKB-to-IDG process in which uncertainty is preserved and resolved progressively rather than collapsed into a single confidence or voting decision.
\end{abstract}


\section{Introduction}
\label{sec:introduction}

Critical infrastructure comprises tightly coupled physical processes,
industrial control systems (ICSs), communication networks, software, and
human operators whose interactions determine how the overall system
functions. These dependencies create paths through which cyber compromises,
component failures, and other disruptions can propagate across interconnected
systems, amplifying their operational impact~\cite{rinaldi2001interdependencies}.
Understanding such relationships is therefore important for cyber-risk
assessment, impact analysis, and resilience evaluation. The knowledge required to model these dependencies is rarely available in a
complete machine-readable form. Instead, it is distributed across technical
and operational documents describing equipment, instrumentation, control
architectures, communication interfaces, process behavior, and operating
procedures~\cite{li2026lsdtsllmaugmentedsemanticdigital}. Knowledge graphs
provide a natural way to organize such information as entities and semantic
relationships~\cite{abu-salih_domain_2021}, but constructing them from
infrastructure evidence remains difficult. Documents contain aliases,
domain-specific terminology, incomplete descriptions, and indirectly
expressed relationships. More importantly, identifying two components does
not establish a dependency between them; the relationship semantics and its
direction must also be determined.

Large language models (LLMs) offer a promising means of automating this
process. Recent work has explored LLM-based knowledge-graph construction and
fusion~\cite{yanggraphfusion2025}, ontology-guided extraction
~\cite{khorshidi2025odkeontologyguidedopendomainknowledge}, evidence-grounded
knowledge construction~\cite{malashin_scigraph_2026}, and broader integration
of LLMs with structured knowledge~\cite{panunifying2024}. LLMs have also been
applied to infrastructure documents for ontology-backed planning
~\cite{li2026lsdtsllmaugmentedsemanticdigital} and combined with existing ICS
knowledge graphs for security analysis~\cite{hosseini_leveraging_2025}.
However, directly converting LLM output into an Infrastructure Dependency
Graph (IDG) is risky: models may omit components, assign inconsistent names
or labels, infer unsupported relationships, or disagree about dependency
meaning and direction. In an IDG, an incorrect endpoint or edge can alter
apparent system connectivity and affect downstream analysis.

Existing approaches address individual parts of this problem, but not the
complete transition from heterogeneous evidence to validated infrastructure
dependencies. Ontology validity does not establish evidence support;
evidence-supported assertions may still refer to the wrong canonical entity;
and cross-model agreement does not establish correctness. Conversely,
requiring majority agreement may discard valid assertions recovered by only
one model.

We address this gap with an evidence-first, multi-LLM framework that
constructs an intermediate \emph{Infrastructure Knowledge Base} (IKB) before
instantiating the final IDG. Multiple open-weight LLMs independently extract
candidate entities and dependencies from the same normalized evidence.
Open-weight models also support local deployment, which is important when
infrastructure documents contain sensitive topology, configuration, and
operational information. Raw extraction precedes ontology grounding, after
which candidates are reconciled with their source evidence, resolved into
canonical entities, aligned to dependency endpoints, validated, and fused
across models. Evidence support, ontology reconciliation, endpoint resolution,
model agreement, and human validation remain separate states throughout the
process. Unresolved cases and provenance are preserved for review, and the
validated IKB is projected deterministically into the IDG without introducing
new LLM-generated assertions.

These design choices motivate four research questions: \textbf{RQ1}, how
evidence-first synthesis affects LLM-generated candidate knowledge;
\textbf{RQ2}, how entity resolution and dependency alignment affect the
transition to canonical infrastructure knowledge; \textbf{RQ3}, how much
cross-model agreement remains after canonicalization and what this implies
for fusion; and \textbf{RQ4}, how closely the automated IKB agrees with a
human-annotated reference representation.

We evaluate the framework across nine critical-infrastructure projects using
three independently executed open-weight LLMs. The results show that evidence
verification and canonical endpoint resolution substantially reduce the set
of dependency assertions, while cross-model overlap is much lower for
dependencies than for entities. Comparison with a human-annotated reference
IKB further shows that entity recovery is substantially stronger than complete
directed dependency recovery. These results identify key bottlenecks in the
evidence-to-graph process.

The principal contributions of this work are:
\begin{itemize}
    \item We formulate infrastructure knowledge construction as an
    evidence-to-IKB-to-IDG process that separates model-generated candidates,
    canonical knowledge, human-validated knowledge, and final graph edges.

    \item We develop an evidence-first multi-LLM architecture that separates
    raw extraction from ontology grounding and verifies candidate knowledge
    against its originating evidence before cross-model reconciliation.

    \item We introduce a provenance-preserving reconciliation process in
    which entity identity precedes dependency fusion, unresolved endpoints
    are retained for review, and model agreement is not used as a majority
    acceptance rule.

    \item We evaluate the framework across nine infrastructure projects and
    characterize how synthesis, entity resolution, endpoint alignment,
    validation, and fusion transform raw LLM assertions into canonical
    infrastructure knowledge.
\end{itemize}

\section{Related Works} 
\noindent\textbf{Critical Infrastructure Dependency Modeling} Critical infrastructures consist of interconnected physical, cyber, and operational components whose dependencies can influence system functionality and failure propagation. Prior studies have modeled such dependencies using graph-based \cite{stergiopoulos2016time, castellanos2018finding, kotzanikolaou2011interdependencies,  hasan2015modeling} dependency-based \cite{rotibi2023extended}, and system-dynamics approaches \cite{genge2015system} to characterize interactions among infrastructure components and assess cascading effects. For example, Canzani et al. \cite{canzani2016characterising} developed a system dynamics model in which infrastructure building blocks are connected through weighted service dependencies to analyze how disruptions propagate across interconnected infrastructures. Adamos et al. \cite{adamos2024enhancing} modeled CPS architectures using state dependency graphs, where directed edges represent logical dependencies between component states, and analyzed dependency paths to assess cascading risks and identify influential components. Our work complements these efforts by constructing system-specific, evidence-supported directional dependencies from heterogeneous infrastructure information before representing the validated relationships in an Infrastructure Dependency Graph.

\noindent\textbf{Knowledge Graph Construction for Industrial Systems} KGs have emerged as a useful approach for integrating heterogeneous industrial information and representing entities and their semantic relationships within a unified structure \cite{shen2020data, kurniawan2024ics}. Prior work has demonstrated the utility of KGs across various industrial tasks, including system modeling \cite{shen2020data}, fault diagnosis \cite{cai2024research}, root cause analysis \cite{martinez2022root}, and operational decision support \cite{chatterjee2022automated}. Existing approaches construct industrial KGs through ontology-driven modeling \cite{yang2023ontology} and integration of structured and unstructured information sources \cite{kumar2022fabkg}. In ICS environments, such approaches have been used to organize system components, cybersecurity concepts, and their relationships into structured representations. However, constructing system-specific infrastructure knowledge remains challenging when relevant information is distributed across heterogeneous industrial data sources \cite{wan2023constructing}. In particular, infrastructure modeling requires not only identifying system entities, but also recovering their directional dependencies and linking those assertions to supporting evidence. This work addresses this challenge by constructing an evidence-grounded Infrastructure Knowledge Base that integrates system entities, directional dependencies, and provenance before validated knowledge is projected into an Infrastructure Dependency Graph.

\noindent\textbf{LLM-Based Knowledge Graph Construction and
Ontology-Guided Extraction.} LLMs are increasingly used to extract entities and
relationships from unstructured text and transform them into knowledge graphs \cite{zhang2024extract,trajanoska2023enhancing,pan2025taxonomy}. They have also been applied to ontology construction,
population, alignment, and entity disambiguation \cite{shimizu2025accelerating}, including in
domain-specific knowledge-graph pipelines \cite{zhou2026llm,van2024ontology,anuyah2025automated}.
Ontology guidance can improve the semantic consistency of
these outputs \cite{schmidt2025document}, but extraction
errors, noise, and hallucinated assertions remain important
limitations \cite{huang2025can}. Ontology conformance establishes whether an assertion can
be represented within a domain model; it does not establish
whether the source evidence supports that assertion. This
distinction is critical for infrastructure modeling, where
an unsupported or incorrectly directed relationship can
produce a misleading dependency edge. Our framework
therefore separates extraction, evidence assessment,
ontology grounding, and cross-model fusion. Multiple LLMs
independently generate candidate knowledge, while evidence
support, ontology reconciliation, and model agreement are
recorded as distinct signals. Assertions retain their
evidence provenance throughout reconciliation and human
validation before entering the validated Infrastructure
Dependency Graph.
\section{Framework Design}
\label{sec:framework}

\begin{figure*}[t]
    \centering
    \includegraphics[width=0.75\textwidth, height= 7.5cm]{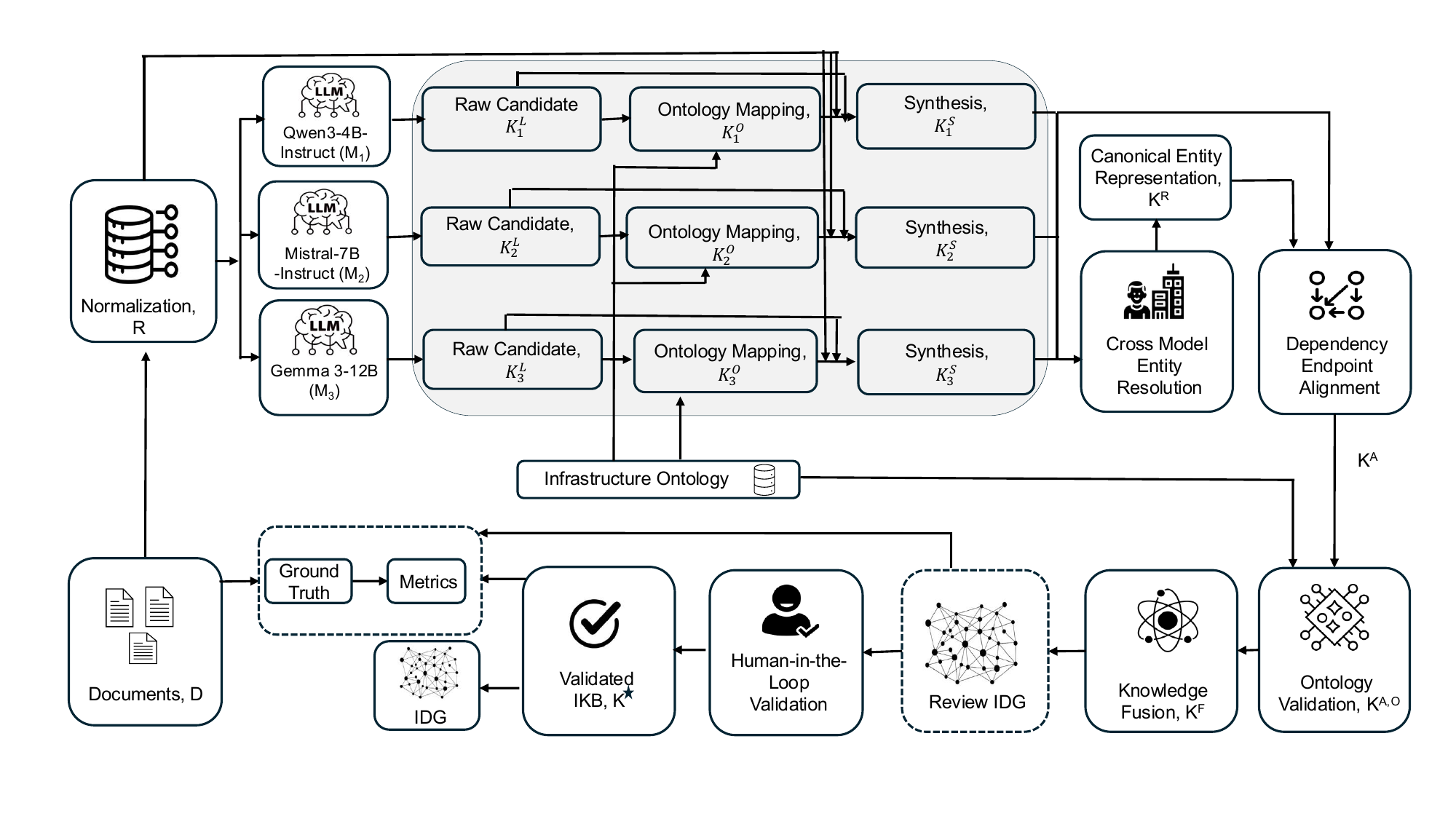}
    \caption{Overview of the evidence-first multi-LLM framework for constructing
an IKB and IDG. Per-model extraction and synthesis are followed by cross-model
resolution, endpoint alignment, ontology validation, fusion, and Human-in-the-Loop
validation. The ground-truth and metrics branch is used only for research
evaluation.}
    \label{fig:framework}
\end{figure*}

The proposed framework transforms heterogeneous critical-infrastructure
evidence into a validated IKB and
subsequently into an IDG. Figure~\ref{fig:framework} provides an overview of the pipeline. The design separates (1) evidence extraction, (2) ontology grounding and evidence-first synthesis, (3) cross-model entity resolution, (4) dependency alignment and ontology validation, (5) knowledge fusion, (6) human-in-the-loop validation, and (7) IDG construction, into distinct stages while preserving provenance throughout the transformation. This separation
is necessary because LLM-generated assertions, ontology-consistent
interpretations, and human-validated knowledge represent different
semantic and validation states and should not be conflated. The overall
transformation is formalized in Equation~\ref{eq:overall_pipeline}.

\begin{align}
\mathcal{D}
&\xrightarrow{\text{CC-CMF}}
\mathcal{R}
\xrightarrow{M_{1:K}}
\mathcal{K}^{L}_{1:K}
\xrightarrow{\mathrm{Map}_{\mathcal{O}}}
\mathcal{K}^{O}_{1:K}
\xrightarrow{\mathrm{Synthesis}}
\mathcal{K}^{S}_{1:K}
\nonumber\\
&\xrightarrow{\text{Entity Resolution}}
\mathcal{K}^{R}
\xrightarrow{\text{Endpoint Alignment}}
\mathcal{K}^{A}
\xrightarrow{\mathrm{Val}_{\mathcal{O}}}
\mathcal{K}^{A,O}
\nonumber\\
&\xrightarrow{\text{Fusion}}
\mathcal{K}^{F}
\xrightarrow{\text{Human Validation}}
\mathcal{K}^{*}
\xrightarrow{\Pi_{\mathrm{IDG}}}
G .
\label{eq:overall_pipeline}
\end{align}

Here, $\mathcal{D}$ denotes the heterogeneous evidence corpus and
$\mathcal{R}$ denotes the normalized evidence records produced through
the Critical Infrastructure Collection Management Framework (CC-CMF). Each model
$M_k$ independently extracts a raw candidate representation
$\mathcal{K}^{L}_{k}$. The extracted terminology is then interpreted
with respect to the infrastructure ontology $\mathcal{O}$, producing
$\mathcal{K}^{O}_{k}$. Rather than allowing ontology mapping to overwrite
the raw model interpretation, an evidence-first synthesis stage reconciles
the raw extraction, its ontology-grounded interpretation, and the
originating evidence to produce $\mathcal{K}^{S}_{k}$. The synthesized
outputs are then reconciled across models. Cross-model entity resolution identifies candidate entities that refer to the same infrastructure component and produces the canonical representation $\mathcal{K}^{R}$. Candidate dependency endpoints are subsequently aligned to the resolved entities, yielding $\mathcal{K}^{A}$. Automated ontology and structural validation produces $\mathcal{K}^{A,O}$, after which eligible assertions are fused across models into $\mathcal{K}^{F}$ while retaining model support, evidence support, disagreement, confidence metadata, and provenance. Human validation constitutes the final decision layer, in which an analyst reviews the synthesized and fused assertions through the framework's interactive interface and decides whether to accept, modify, reject, merge, or override each canonical entity and dependency. Entity decisions are applied before final dependency decisions so that a dependency cannot be retained when either canonical endpoint has been rejected or otherwise made inactive. The resulting validated IKB $\mathcal{K}^{*}$ is then projected into the IDG $G$ according to the ontology and the selected graph export policy.

\subsection{Input Evidence Model}
\label{sec:input_evidence}

Critical-infrastructure knowledge is rarely contained in a single source.
Information about assets, processes, and dependencies is distributed across
heterogeneous technical artifacts and human knowledge. For example, an
operator interview may state that \emph{Pump P-101 requires manual restart
following a PLC failure}, while an operating procedure may independently
describe the same dependency. Likewise, a controller may appear in an
equipment inventory, a network diagram, and a control narrative. Constructing
an IKB therefore requires integrating heterogeneous evidence while preserving
its origin.

To support this requirement, we use CC-CMF, a structured methodology for
collecting and organizing infrastructure evidence. Sources may include
engineering documents, asset inventories, network and process diagrams,
configurations, control logic, operating and maintenance procedures,
organizational records, interviews, walkthroughs, and observations. CC-CMF
does not assume that any single source is complete; instead, sources may
agree, supplement, or contradict one another.

Let $\mathcal{P}$ denote the set of infrastructure environments. For each
project $p\in\mathcal{P}$, the collected evidence corpus is

\begin{equation}
\mathcal{D}_{p}
=
\{d_{p,1},d_{p,2},\ldots,d_{p,M_p}\},
\label{eq:evidence_corpus}
\end{equation}

where $M_p$ is the number of source artifacts associated with that
environment. CC-CMF converts the collected corpus into a common evidence
representation,

\begin{equation}
\mathcal{R}
=
\Gamma_{\mathrm{CC\text{-}CMF}}(\mathcal{D})
=
\{r_1,r_2,\ldots,r_N\},
\label{eq:evidence_normalization}
\end{equation}

where $\Gamma_{\mathrm{CC\text{-}CMF}}(\cdot)$ denotes evidence registration,
segmentation, and normalization. An evidence record may correspond to a
document paragraph, table row, configuration block, diagram element,
interview statement, or recorded observation.

Each normalized record is represented as

\begin{equation}
r_n = \langle x_n,s_n,\ell_n,\pi_n,p_n\rangle,
\label{eq:evidence_record}
\end{equation}

where $x_n$ is the evidence content, $s_n$ its structural form, $\ell_n$ its
location in the source, $\pi_n$ its provenance metadata, and $p_n$ the
associated infrastructure project. Project identity is retained so that
subsequent entity resolution and dependency alignment remain project-scoped.

An evidence record is not itself an infrastructure assertion; it is the
source from which candidate assertions are extracted. Multiple records may
independently support the same entity or binary dependency, while agreement
among multiple LLMs over the same record represents multiple model
interpretations rather than independent evidence. This distinction allows
the framework to track \emph{evidence support} separately from
\emph{model agreement}. Normalization therefore introduces no entities,
dependencies, or ontology assignments; it produces only a common,
traceable evidence representation for subsequent multi-LLM extraction.

\subsection{Infrastructure Knowledge Base Representation}
\label{sec:ikb_representation}

We represent the Infrastructure Knowledge Base as
\begin{equation}
\mathcal{K}
=
\left(
\mathcal{V},\mathcal{A},\mathcal{E},\mathcal{P}
\right),
\label{eq:ikb}
\end{equation}
where $\mathcal{V}$ contains canonical infrastructure entities,
$\mathcal{A}$ their attributes, $\mathcal{E}$ dependency assertions, and
$\mathcal{P}$ provenance linking entities and dependencies to supporting
evidence. During construction, unresolved, conflicting, or review-required
assertions are retained separately in a temporary state $\mathcal{U}$.

Each canonical entity $v_i$ has a unique identifier, project identifier,
canonical name, ontology class, infrastructure layer, validation state,
and optional attributes and confidence information. Entities are assigned
to one of five layers: cyber ($L_C$), operational ($L_O$), physical
($L_P$), human ($L_H$), or organizational ($L_G$).

A dependency is represented as
\begin{equation}
e_m=(p_m,v_s,v_t,\rho_m,\tau_m),
\end{equation}
where $p_m$ identifies the project, $v_s$ and $v_t$ are the source and
target entities, $\rho_m$ is the relationship, and $\tau_m$ is the broader
dependency type. The direction $v_s\rightarrow v_t$ means that $v_t$
depends on, or is materially supported by, $v_s$.

An \emph{endpoint} is the role an entity occupies as the source or target
of a dependency, rather than a separate object type. During extraction,
an endpoint may initially be a raw model-generated entity mention; it
becomes a \emph{canonical endpoint} after resolution to an entity
$v_i\in\mathcal{V}$. A canonical dependency requires both endpoints to
resolve within the same infrastructure project.

The relationship $\rho_m$ and dependency type $\tau_m$ serve different
purposes. The relationship captures the specific semantic connection and
is part of the canonical dependency identity,
\begin{equation}
(p_m,v_s,v_t,\rho_m).
\end{equation}
The dependency type is classification metadata and may remain unresolved
when the relationship itself is retained. The ontology defines nine types:
\textsc{Control}, \textsc{Data}, \textsc{Communication},
\textsc{Operational}, \textsc{Physical}, \textsc{Human},
\textsc{Organizational}, \textsc{Resource}, and \textsc{Recovery}.

For any entity or dependency $a$, $\mathcal{P}(a)$ preserves its supporting
evidence and contributing model assertions. Provenance is accumulated during
resolution and fusion rather than replaced, allowing each canonical object
to remain traceable to its source evidence.

\subsection{Multi-LLM Candidate Knowledge Extraction}
\label{sec:multi_llm_extraction}

Given normalized evidence, open-weight models
$\mathcal{M}=\{M_1,\ldots,M_K\}$ independently construct
model-specific candidate knowledge using the same
instructions and output schema. Extraction is
ontology-independent: source terminology is preserved, and
ontology grounding occurs only in the next stage.

Normalized records from the same source artifact are grouped
into extraction units $X_q\subseteq\mathcal{R}$ to bound
context while preserving provenance. Each record has a
unique identifier that extracted assertions must cite.

Extraction uses two passes:

\begin{equation}
\begin{aligned}
\mathcal{V}^{L}_{k,q}
  &= M_k(X_q,\Pi_V),\\
\mathcal{E}^{L}_{k,q}
  &= M_k(X_q,\mathcal{V}^{L}_{k,q},\Pi_E),
\end{aligned}
\label{eq:two_pass_extraction}
\end{equation}

where $\Pi_V$ and $\Pi_E$ denote the entity and dependency
instructions. The first pass retains raw entity names,
descriptions, attributes, and model-proposed types. The
second uses these candidates to promote consistent endpoint
naming while keeping entity identification separate from
dependency reasoning. Previously missed endpoints may still
be proposed, but they must later resolve to canonical
entities. Raw relationship phrases, optional free-text
dependency types, and direction are retained for synthesis.

Aggregating all units produces
$\mathcal{K}^{L}_{k}=(\mathcal{V}^{L}_{k},
\mathcal{A}^{L}_{k},\mathcal{E}^{L}_{k},
\mathcal{P}^{L}_{k})$, where $\mathcal{P}^{L}_{k}$ preserves
source-record references. Assertions with invalid references
are discarded. Model outputs remain independent at this
stage, without ontology correction, canonicalization, or
cross-model fusion.

\subsection{Ontology Grounding and Evidence-First Synthesis}
\label{sec:ontology_synthesis}

\noindent\textbf{Ontology grounding.}
Following raw extraction, entity types, relationships, and
dependency types are mapped to the infrastructure ontology.
Deterministic normalized matching is applied first, followed
by an ontology-constrained LLM mapper for unresolved terms.
The mapper selects only from the supplied vocabulary and
classifies each result as \textsc{Direct-Equivalent},
\textsc{Safe-Superclass}, \textsc{Related-Only}, or
\textsc{Unmapped}. Entity types may use \textsc{Direct-Equivalent} or
\textsc{Safe-Superclass} mappings, whereas relationships and
dependency types require \textsc{Direct-Equivalent} for
automatic use. \textsc{Related-Only} is retained for review,
and \textsc{Unmapped} indicates that no defensible mapping
was found. The resulting $\mathcal{K}^{O}_{k}$ augments
rather than replaces $\mathcal{K}^{L}_{k}$; ontology mapping
does not provide source evidence or additional model
agreement.

\noindent\textbf{Evidence-first synthesis.}
Because ontology consistency does not establish factual
support, the framework reconciles each raw candidate, its
ontology interpretation, and its originating evidence:

\begin{equation}
\mathcal{K}^{S}_{k}
=
\operatorname{Synth}
\left(
\mathcal{K}^{L}_{k},
\mathcal{K}^{O}_{k},
\mathcal{R}
\right).
\label{eq:evidence_first_synthesis}
\end{equation}

Synthesis first determines whether the cited evidence
supports the candidate and then reconciles its
evidence-supported meaning with the ontology. Entity
evidence is classified as \textsc{Supported},
\textsc{Unsupported}, or \textsc{Ambiguous}. A supported
entity may retain its mapped class (\textsc{Accept}), receive
a more faithful available class (\textsc{Revise}), or remain
an \textsc{Ontology-Gap} when the ontology has no adequate
class.

Dependency synthesis applies a stricter gate and records the
five dimensions summarized in
Table~\ref{tab:dependency_axes}. Only a relationship with
supported evidence and consistent direction proceeds along
the automatic reconciliation path; ontology mapping cannot
rescue an unsupported, ambiguous, or reversed assertion.
Relationship and dependency type are reconciled separately.
Dependency-type evidence permits \textsc{Not-Stated}
because a relationship may be supported without an explicit
broader dependency class. A missing or unresolved dependency
type does not invalidate an otherwise admissible
relationship.

\begin{table}[t]
\centering
\caption{Classification dimensions tracked during
dependency synthesis.}
\label{tab:dependency_axes}
\small
\setlength{\tabcolsep}{3pt}
\begin{tabular}{@{}
    p{0.20\columnwidth}
    p{0.25\columnwidth}
    p{0.47\columnwidth}
@{}}
\toprule
\textbf{Object} & \textbf{Dimension} & \textbf{Outcomes} \\
\midrule

Relationship
    & Evidence support
    & Supported, Unsupported, Ambiguous \\

Relationship
    & Direction
    & Consistent, Reversed, Ambiguous \\

Relationship
    & Reconciliation status
    & Accept, Revise, Ontology-Gap, Unresolved, Conflict \\

Dependency type
    & Evidence support
    & Supported, Unsupported, Ambiguous, Not-Stated \\

Dependency type
    & Reconciliation status
    & Accept, Revise, Ontology-Gap, Unresolved, Conflict \\

\bottomrule
\end{tabular}
\end{table}

Synthesis preserves evidence support separately from
ontology reconciliation. Assertions that do not pass the
evidence-first gate remain outside the automatic downstream
path but are retained for review or later resolution.

\subsection{Cross-Model Entity Resolution}
\label{sec:entity_resolution}

After synthesis, entity candidates from different models are
resolved into project-scoped canonical identities before
dependency alignment. For two candidates $v_i$ and $v_j$,
the resolver considers normalized-name similarity,
deterministic-identifier agreement, attribute agreement, and
synthesized-type agreement. Let $\mathcal{S}_{ij}$ be the
features available for both candidates. The resolution score
is

\begin{equation}
S_{\mathrm{res}}(v_i,v_j)
=
\frac{\sum_{r\in\mathcal{S}_{ij}}
\alpha_r\,\mathrm{sim}_r(v_i,v_j)}
{\sum_{r\in\mathcal{S}_{ij}}\alpha_r},
\label{eq:entity_resolution_score}
\end{equation}

where unavailable features are excluded rather than treated
as disagreement.

Deterministic identifiers, including equipment tags, serial
numbers, network addresses, and hostnames, provide stronger
identity evidence than name similarity. A shared identifier can establish a
match despite name variation, whereas conflicting
identifiers prevent automatic merging. Lexical similarity
alone cannot establish identity.

Each pair is classified as \textsc{Auto-Match},
\textsc{Review}, or \textsc{Separate}. Only
\textsc{Auto-Match} pairs participate in automatic
clustering, while \textsc{Review} cases remain distinct
until explicitly resolved. The candidate-to-canonical
mapping is preserved for dependency alignment.

\noindent\textbf{Attribute reconciliation.}
Attributes within each canonical cluster are normalized and
grouped by key and value. The most frequent normalized value
is retained provisionally, while alternatives and their
model, record, and evidence provenance are preserved.
Conflicting values are not discarded, and numerical values
are not averaged.

\subsection{Dependency Alignment and Ontology Validation}
\label{sec:dependency_alignment}

Dependencies are reconciled only after entity candidates have been resolved
to canonical identities. This ensures that relationship assertions are not
compared or fused across models until their source and target endpoints refer
to the same canonical entities.

\noindent\textbf{Endpoint alignment.}
A synthesized dependency
$e=(p,s,t,\rho,\tau,\pi)$ retains its raw source and target mentions $s$ and
$t$, together with its relationship, dependency type, and provenance.
Successful alignment produces

\begin{equation}
e^{A}
=
\left(
p,
\alpha(s),
\alpha(t),
\rho,
\tau,
\pi
\right),
\label{eq:aligned_dependency}
\end{equation}

where $\alpha(\cdot)$ maps a raw endpoint mention to a canonical entity
within the same project.

Alignment is deterministic and conservative. Canonical names and exact
aliases established during entity resolution are used, with same-project,
same-model, and same-record matches preferred when available. Cross-project
alignment and fuzzy endpoint assignment are not permitted. If an endpoint
cannot be resolved, the dependency remains unresolved; ambiguous matches are
retained for review rather than assigned arbitrarily. Alignment changes only
endpoint identity, while the synthesized relationship, dependency type, and
provenance are preserved.

\noindent\textbf{Ontology and structural validation.}
Once both endpoints are canonicalized, the aligned assertion is checked
against the active ontology $\mathcal{O}$ and structural constraints. This
stage determines whether the assertion is admissible for the canonical IKB;
factual support is not reassessed because it has already been established
during evidence-first synthesis. Entity validation checks that the synthesized entity type is represented in
the ontology and that the entity has an admissible infrastructure layer.
Dependency validation requires both endpoints to exist within the same
project and the relationship to correspond to an ontology-defined dependency
relation eligible for IDG construction. Descriptive relationships are
therefore excluded from IDG edges. Relationship validity and dependency-type validity are evaluated separately.
An unresolved dependency type does not invalidate an otherwise admissible
relationship; when a type is assigned, it must belong to the ontology
vocabulary and be compatible with the relationship. The resulting representation is denoted $\mathcal{K}^{A,O}$.
Ontology-valid assertions enter the automatic fusion path, while unresolved
endpoints, ontology gaps, structural violations, and incompatible
classifications remain available for review. Automated ontology validation
remains distinct from human acceptance.

\subsection{Knowledge Fusion}
\label{sec:knowledge_fusion}

Knowledge fusion consolidates model support and provenance after entity
canonicalization, dependency alignment, and automated validation. Canonical
assertions do not require majority support: a valid assertion recovered by
only one model may survive, while support from additional models is retained
as corroborating metadata. For entities, fusion preserves the canonical identity established during
entity resolution and aggregates the contributing model assertions,
evidence records, source locations, name and type variants, attributes, and
validation metadata. Distinct-model agreement is computed from the number of
independently executed models contributing to the canonical entity; repeated
assertions from the same model do not increase this support. Dependency fusion operates over aligned, ontology-valid assertions. A canonical dependency is identified by

\begin{equation}
\sigma(e)
=
\left(
p,\hat{v}_s,\hat{v}_t,\rho
\right),
\label{eq:canonical_dependency_signature}
\end{equation}

where $p$ is the project, $\hat{v}_s$ and $\hat{v}_t$ are canonical
endpoints, and $\rho$ is the synthesis-verified relationship. Assertions
sharing this signature are consolidated while preserving their individual
model and evidence provenance. Dependency type $\tau$ is excluded from the
canonical identity because it is classification metadata rather than the
relationship itself.

If all contributing assertions agree on $\tau$, that type is retained.
When competing types are proposed, distinct-model support is compared; a
unique highest-supported type is selected, while a tie leaves the type
unresolved and records an explicit \textsc{Conflict}. The underlying
relationship is retained. Different relationships between the same canonical
endpoints are likewise preserved as distinct dependencies rather than
resolved by majority voting.

After fusion, confidence is computed separately as

\begin{equation}
c(e)
=
\gamma_M A_M(e)
+
\gamma_E Q_E(e)
+
\gamma_O Q_O(e)
+
\gamma_R Q_R(e),
\label{eq:fusion_confidence}
\end{equation}

where $A_M(e)$ measures distinct-model agreement, $Q_E(e)$ support from
distinct normalized evidence records, $Q_O(e)$ ontology consistency, and
$Q_R(e)$ source reliability. Model agreement and evidence support remain
separate: several models interpreting the same evidence record increase
model agreement but do not create additional independent evidence.
Confidence is retained as supporting metadata and does not determine final
acceptance into the validated IKB.

\subsection{Human-in-the-Loop Validation}
\label{sec:human_validationdation_idg}

Automated grounding, resolution, validation, and fusion do
not determine whether a \textbf{canonical assertion}
should become final infrastructure knowledge. The framework
therefore applies a human validation gate before constructing
the validated IKB and IDG, keeping ontology conformance
separate from human acceptance.

\noindent\textbf{Entity validation.}
Review proceeds entity-first. Each canonical entity is
presented with its synthesized type, layer, attributes,
evidence, model support, confidence, conflicts, and
provenance. A reviewer may accept, modify, reject, merge,
override, or leave the entity unresolved. Entity decisions
are propagated to affected dependencies. Merges rewrite
dependency endpoints to the surviving entity and reopen
prior dependency decisions when necessary. A dependency can
receive a final decision only when both endpoints are active
and human-retained; this is the \emph{entity-first trust
gate}.

\noindent\textbf{Dependency validation.}
After endpoint reconciliation, reviewers inspect each
dependency with its endpoints, relationship, type, evidence,
model support, ontology status, confidence, conflicts, and
provenance. The same review actions are available, and
unresolved or conflicting dependency types remain explicit. An ontology override may retain an evidence-supported
assertion not fully represented by the ontology, but it
cannot bypass structural constraints. Both endpoints must
remain active, human-retained, and within the same project,
and the relationship must be eligible for IDG export.
Overrides are explicitly marked as human-certified
exceptions. The resulting validated IKB,
$\mathcal{K}^{*}=(\mathcal{V}^{*},\mathcal{A}^{*},
\mathcal{E}^{*},\mathcal{P}^{*})$, contains human-retained
entities, attributes, and dependencies that satisfy the
entity-first trust gate while preserving their accumulated
evidence, model, fusion, and review provenance.

\subsection{IDG Construction}
\label{sec:idg}
The IDG is constructed deterministically from the validated IKB:
\begin{equation}
G
=
\Pi_{\mathrm{IDG},\mathcal{O}}
\left(
\mathcal{K}^{*}
\right)
=
(V_G,E_G),
\label{eq:idg_projection}
\end{equation}
where $V_G$ and $E_G$ denote the IDG node and edge sets. Each retained
canonical entity becomes a graph node. A validated dependency becomes a
directed edge only when both canonical endpoints are present in the
graph, belong to the same project as the dependency, and its relationship
is designated by the ontology as exportable to the IDG. Relationships
that are descriptive or otherwise not admissible as IDG edges are
excluded during projection. Distinct relationships between the same source and target are preserved,
yielding a directed multigraph. Canonical dependency identifiers are used
as edge identifiers, allowing multiple dependency relationships to
coexist between the same endpoint pair. Nodes and edges retain canonical
identifiers and validation metadata, preserving a path back through the
validated IKB to their evidence and model provenance. Graph construction
therefore introduces no new LLM-generated assertions; it deterministically
projects eligible human-validated infrastructure knowledge under the
ontology's IDG export policy.

\section{Implementation}
\label{sec:implementation}

We implement the framework in Python as a local,
provenance-preserving pipeline. SQLite and SQLAlchemy store
evidence, model outputs, ontology mappings, synthesis
decisions, canonical assertions, conflicts, and review
actions. Streamlit supports project-scoped inspection and
human validation, and NetworkX represents each IDG as a
directed multigraph.

\noindent\textbf{Evidence and Model Configuration.}
The runner processes one registered evidence artifact per
project and divides it into project-scoped records and
extraction units. Each record retains its source location
and identifier, allowing subsequent assertions to be traced
to the supporting evidence. Three open-weight models
independently process the same units:
Qwen3-4B-Instruct~\cite{yang2025qwen3technicalreport},
Gemma 3-12B~\cite{gemmateam2025gemma3technicalreport}, and
Mistral-7B-Instruct~\cite{jiang2023mistral7b}, with 4.0B,
12.2B, and 7.2B parameters, respectively. All models run
locally through Ollama. Qwen and Gemma use Q4\_K\_M
quantization, while Mistral uses Q4\_0. We use prompt
version v16, temperature 0.0, top-$p=1.0$, seed 42, an
8,192-token output limit, and a 16,384-token context.
Model digests, prompt and evidence hashes, project
identifiers, and run identifiers are recorded for
reproducibility.

\noindent\textbf{Grounding, Resolution, and Fusion.}
The implementation uses version 0.1.1 of the CC-CMF
critical-infrastructure ontology. Grounding begins with
deterministic normalized matching. Qwen3-4B-Instruct maps
unmatched terms using an ontology-constrained prompt, and
eligible mappings are promoted only when confidence is at
least 0.80. Evidence-first synthesis then evaluates each
candidate against its cited evidence and ontology
interpretation. Raw extractions, mapping decisions, and
synthesis outcomes remain separate to preserve provenance. Entity resolution is restricted to candidates from the same
project. Its score combines name, identifier, attribute,
and synthesized-type agreement with weights of 0.30, 0.40,
0.20, and 0.10. Scores of at least 0.85 are eligible for
automatic matching, while scores in $[0.60,0.85)$ are
retained for review. Name similarity alone cannot establish
identity. Dependency endpoints are aligned to same-project
canonical names and aliases, with fuzzy assignment
disabled. Ontology-valid assertions sharing a project,
canonical endpoints, and synthesized relationship are
fused while retaining their evidence and model provenance.
Dependency type is reconciled separately and does not
determine canonical dependency identity. Dependency
confidence combines model agreement, evidence support,
ontology consistency, and source reliability with weights
of 0.30, 0.25, 0.20, and 0.25; it supports review
prioritization but does not determine acceptance.

\noindent\textbf{Human Validation and IDG Export.}
The review interface presents each canonical assertion with
its evidence, model support, ontology status, conflicts,
confidence, alternatives, and provenance. Review proceeds
entity-first: entity decisions are propagated to affected
dependency endpoints, and a dependency can be retained only
when both endpoints are active and human-retained. Review
actions are recorded without replacing the automated
provenance. The review IDG contains assertions under
inspection, whereas the validated IDG contains only
human-retained assertions satisfying the endpoint and
export rules. IDG construction is deterministic and
introduces no additional LLM-generated assertions.
\section{Evaluation}
\label{sec:evaluation}

We evaluate the canonical pre-human IKB produced after
grounding, synthesis, resolution, alignment, and fusion.
Following the research questions in
Section~\ref{sec:introduction}, we examine synthesis
outcomes (RQ1), canonicalization and alignment (RQ2),
cross-model support (RQ3), and agreement with a
human-annotated reference IKB (RQ4). Raw candidates are used
only to characterize the earlier pipeline stages.

\noindent\textbf{Corpus and Reference IKB.}
The evaluation covers nine natural-gas, water/wastewater,
and electric-distribution projects, with one evidence
document per project. The corpus contains 702 normalized
records and 356 extraction units. The reference IKB contains
284 entities and 102 directed dependencies
(Table~\ref{tab:evaluation_corpus}).

\begin{table*}[t]
\centering
\caption{Evaluation corpus and human-annotated reference IKB.
Records/Units denotes normalized records and extraction
units; E/D denotes reference entities and dependencies.}
\label{tab:evaluation_corpus}
\small
\setlength{\tabcolsep}{7pt}
\begin{tabular}{lllrr}
\toprule
\textbf{Project} &
\textbf{Sector} &
\textbf{Evidence Artifact} &
\textbf{Records/Units} &
\textbf{Reference E/D} \\
\midrule
AusNet Gas
  & Natural gas distribution
  & SCADA strategy
  & 36/21
  & 19/7 \\

Boys Water
  & Water/wastewater
  & Synthetic company-system design
  & 6/4
  & 31/10 \\

Iqaluit
  & Water/wastewater
  & SCADA system user manual
  & 234/128
  & 30/10 \\

MANK2
  & Electric distribution
  & Synthetic company-system design
  & 10/9
  & 48/14 \\

Parris Island
  & Wastewater
  & SCADA upgrade report
  & 26/14
  & 16/8 \\

Pyramid NG
  & Natural gas midstream
  & Synthetic infrastructure company profile
  & 11/11
  & 40/20 \\

Snoqualmie
  & Water/wastewater
  & SCADA master plan
  & 68/58
  & 22/14 \\

TigerSkid
  & Natural gas compression
  & Control-system platform description
  & 300/100
  & 54/12 \\

Washington IL
  & Water/wastewater
  & SCADA master plan
  & 11/11
  & 24/7 \\
\midrule
\multicolumn{3}{l}{\textbf{Total}}
  & \textbf{702/356}
  & \textbf{284/102} \\
\bottomrule
\end{tabular}
\end{table*}

\noindent\textbf{Human Annotation.}
We manually reviewed each document to construct the
reference IKB. For entities, we recorded canonical names,
aliases, ontology types, and infrastructure layers. For
dependencies, we identified the endpoints, direction,
relationship, and dependency type. We included only
assertions with clear documentary support and excluded
ambiguous cases. The resulting reference is conservative:
an unmatched prediction is counted as a false positive but
is not necessarily factually incorrect.

\noindent\textbf{Evaluation Representation.}
We evaluate canonical pre-human assertions and count each
exact duplicate normalized identity once per project. Raw
extractions, mappings, and synthesis results remain as
provenance but are not counted as separate predictions.

\noindent\textbf{Entity Matching.}
Names are matched within project after case folding,
trimming, and collapsing repeated whitespace. A prediction
matches only if it equals a reference canonical name or
declared alias. This normalization does not remove
punctuation, join compound words, normalize hyphenation, or
reconcile singular and plural forms. Ontology type and
infrastructure layer are evaluated separately after
identity matching. For example,
\textit{bypass valve (BPV-1)} matches \textit{BPV-1}
because it is a declared alias.

\noindent\textbf{Dependency Matching.}
Endpoint agreement requires both canonical entities to
match in the correct direction. Complete relationship
agreement additionally requires the
source--relationship--target tuple to match. Dependency type
is evaluated separately after the complete relationship is
recovered.

\noindent\textbf{Measures.}
RQ1 reports candidate-level evidence and direction
outcomes; RQ2 reports changes introduced by resolution,
alignment, ontology validation, and fusion; and RQ3 reports
distinct-model support. These are pipeline diagnostics
rather than correctness measures. For RQ4, we pool true
positives, false positives, and false negatives across
projects and report micro-averaged precision, recall, and
$F_1$. Ontology type and layer are evaluated conditionally
for matched entities, and dependency type is evaluated for
complete relationship matches. To diagnose low entity precision, we group unmatched
identities by project, ontology type, and number of
contributing candidates. We also apply broader diagnostic
normalization of punctuation, hyphenation, compound-word
segmentation, and singular/plural forms. This diagnostic
analysis does not change the primary exact-match scores.

\subsection{Results}\label{sec:results}

We report the results in pipeline order. RQ1--RQ3
characterize how candidate assertions are synthesized,
canonicalized, and fused. RQ4 compares the resulting
canonical IKB with Ground Truth.

\subsubsection{RQ1: Evidence-First Synthesis Outcomes}

Table~\ref{tab:extraction_synthesis_v2} summarizes the raw
candidate assertions produced across the nine projects.
The three models generated 5,269 entity candidates and
3,215 dependency candidates.

\begin{table}[t]
\centering
\caption{Candidate generation and evidence-first synthesis
outcomes across nine projects. Parentheses report
class-specific percentages; dashes indicate that an outcome
does not apply.}
\label{tab:extraction_synthesis_v2}
\footnotesize
\setlength{\tabcolsep}{5pt}
\renewcommand{\arraystretch}{1.08}

\begin{tabular}{@{}lrr@{}}
\toprule
\textbf{Model or Outcome} &
\textbf{Entities} &
\textbf{Dependencies} \\
\midrule

\multicolumn{3}{@{}l}{\textit{A. Raw extraction by model}} \\
\addlinespace[2pt]
Qwen3-4B     & 1,607 & 850   \\
Gemma3-12B   & 1,448 & 758   \\
Mistral-7B   & 2,214 & 1,607 \\
\cmidrule(lr){1-3}
\textbf{Total} & \textbf{5,269} & \textbf{3,215} \\

\addlinespace[5pt]
\multicolumn{3}{@{}l}{\textit{B. Evidence assessment}} \\
\addlinespace[2pt]
Supported
  & 5,262 (99.9\%)
  & 1,504 (46.8\%) \\

Ambiguous
  & --
  & 1,234 (38.4\%) \\

Unsupported
  & 7 (0.1\%)
  & 477 (14.8\%) \\

\addlinespace[5pt]
\multicolumn{3}{@{}l}{\textit{C. Dependency direction}} \\
\addlinespace[2pt]
Consistent
  & --
  & 1,504 (46.8\%) \\

Ambiguous
  & --
  & 1,652 (51.4\%) \\

Reversed
  & --
  & 59 (1.8\%) \\
\bottomrule
\end{tabular}
\end{table}

Evidence-first synthesis classified 5,262 entity candidates
(99.9\%) as \textsc{Supported} and seven (0.1\%) as
\textsc{Unsupported}. Synthesis therefore had little
filtering effect on entities, although it could still revise
their ontology interpretation. Dependencies were substantially more difficult to verify.
Of the 3,215 dependency candidates, 1,504 (46.8\%) were
classified as \textsc{Supported}, 1,234 (38.4\%) as
\textsc{Ambiguous}, and 477 (14.8\%) as
\textsc{Unsupported}. Direction was evaluated independently
over the same candidates. Synthesis classified 1,504
directions (46.8\%) as \textsc{Consistent}, 1,652 (51.4\%)
as \textsc{Ambiguous}, and 59 (1.8\%) as
\textsc{Reversed}. These results show that identifying entity mentions was
usually well supported by the cited evidence. Establishing
a directed interaction between two entities was much more
difficult. Evidence-first synthesis therefore acted as a
stronger trust boundary for dependencies than for entities.

\subsubsection{RQ2: Canonicalization and Dependency Alignment}

The 5,262 evidence-supported entity candidates produced
4,068 canonical entity records after project-scoped entity
resolution. This represents a reduction of 1,194 records,
or 22.7\%. Candidates that could not be merged safely were
retained as separate canonical records for human inspection
rather than discarded. Dependency construction was more restrictive. Of the 1,504
evidence-supported dependency candidates, two were routed
as synthesis conflicts and 1,502 entered endpoint
alignment. Of these, 403 (26.8\%) were aligned
automatically, 576 (38.3\%) required review, and 523
(34.8\%) remained unresolved. Ontology validation retained
231 of the 403 aligned assertions. Fusion consolidated
these 231 assertions into 221 canonical dependencies. These results show that documentary support alone is
insufficient to construct a canonical dependency. Both
endpoints must resolve to canonical entities in the correct
direction, and the resulting assertion must satisfy the
ontology and structural checks required for fusion.

\subsubsection{RQ3: Cross-Model Support}

Cross-model support was substantially higher for entities
than for dependencies. Of the 4,068 canonical entity
records, 873 (21.5\%) were supported by multiple models,
whereas 3,195 (78.5\%) had single-model support. By
contrast, only six of the 221 canonical dependencies
(2.7\%) had multi-model support; five were supported by two
models and one by all three. The remaining 215 dependencies
(97.3\%) had single-model support. These results show that the models recovered largely
different dependencies even after canonicalization.
Requiring majority agreement would therefore eliminate most
of the dependency knowledge available for review. The
framework instead retains evidence-supported single-model
assertions and records model agreement as provenance and
review-prioritization metadata.

\subsubsection{RQ4: Agreement with the Human Reference IKB}

Table~\ref{tab:primary_accuracy} compares the canonical
pre-human IKB with the human-annotated reference IKB. The
evaluation collapses exact duplicate identities within each
project, reducing 4,068 stored entity records to 2,692
unique predictions. Similarly, the 221 stored dependencies
produce 218 unique source--relationship--target predictions. The pipeline recovered 197 of the 284 reference entities.
Recall was therefore substantially higher than precision,
as shown in Table~\ref{tab:primary_accuracy}. This imbalance
reflects the role of the pre-human IKB as a broad review
set: all canonical identities are evaluated before a
reviewer determines which should enter the validated IKB.
Exact duplicate identities do not explain the low precision
because they are removed before scoring.

The diagnostic analysis identifies several contributing
factors. \textsc{Iqaluit}, \textsc{Snoqualmie}, and
\textsc{TigerSkid} contain 80.3\% of the extraction units
and account for 73.3\% of unmatched entity identities.
Nearly two thirds of the unmatched identities are derived
from only one contributing candidate assertion, showing
that one-off extractions form a large part of the review
set. The output is also considerably more fine-grained than
the reference: 686 unmatched identities are PLC,
remote-I/O, or HMI items, compared with 38 such entities in
the reference. Strict identity matching contributes a smaller but
measurable source of disagreement. We identified 91
unmatched predictions that differ from a reference name
only through punctuation, spacing, or simple
singular/plural form. Examples include
\textit{waste water treatment plant} versus
\textit{wastewater treatment plant} and
\textit{distribution transformer} versus
\textit{distribution transformers}. The reported precision
therefore measures exact agreement between a broad
pre-human review set and a conservative reference. It
should not be interpreted as the proportion of assertions
that are factually incorrect. Semantic agreement was stronger once entity identity had
been recovered. Ontology type was correct for 75.7\% of
matched entities with an available reference type, while
infrastructure layer was correct for 70.6\% of matched
entities. Entity recovery and consolidation therefore
presented a greater challenge than classifying an already
matched entity.

\begin{table}[t]
\centering
\caption{Micro-averaged agreement between the automated
pre-human IKB and the human-annotated reference IKB across
nine projects.}
\label{tab:primary_accuracy}
\small
\setlength{\tabcolsep}{3.5pt}
\begin{tabular}{@{}lrrr@{}}
\toprule
\textbf{Metric} &
\textbf{Precision (\%)} &
\textbf{Recall (\%)} &
\(\boldsymbol{F_1}\) \textbf{(\%)} \\
\midrule
Entity identity
  & 7.32 & 69.37 & 13.24 \\
Dependency endpoints
  & 1.86 & 3.92 & 2.52 \\
Dependency relationship
  & 0.46 & 0.98 & 0.63 \\
\bottomrule
\end{tabular}
\end{table}

Dependency agreement was constrained primarily by endpoint
identity. Of the 215 unique predicted endpoint pairs, 150
contained no reference-matched endpoint and 45 contained
only one; thus, only 20 connected two individually matched
entities. Four of these 20 pairs recovered a reference edge
in the correct direction. Of the remaining 16, two reversed
a reference edge and 14 did not correspond to a reference
pair. Relationship classification reduced agreement
further: three of the four correctly directed edges had a
different relationship, leaving one complete relationship
and dependency-type match. Although this sample is too
small to assess dependency-type classification, it shows
that most disagreement arose during entity and endpoint
recovery. Project-level results in
Appendix~\ref{app:project_results} show substantial
variation across the evidence collections. Overall, entity consolidation and endpoint resolution remain the principal
barriers to automatic IDG construction. The pre human-in-the-loop validated IKB nevertheless retains unmatched assertions and their provenance for human review. Measuring how this review changes accuracy and analyst effort remains future work.

\subsubsection{Human-in-the-Loop Validation}
\label{sec:human_validation_status}

RQ4 evaluates the automated IKB before human validation,
allowing its performance to be measured independently of
reviewer intervention. The framework nevertheless supports
interactive review, through which analysts can accept,
modify, reject, merge, or override canonical assertions
before they enter the validated IKB and IDG. Evaluating how
this process changes accuracy and the effort required from
reviewers is left to future work.

\section{Discussion}
\label{sec:discussion}

The evaluation shows how the framework transforms
LLM-generated candidates into a provenance-preserving,
human-reviewable IKB and identifies the main limits of this
process.

\noindent\textbf{Why evidence and ontology checks must remain
separate (RQ1).}
Entity candidates were usually supported by their evidence,
whereas dependencies were more often ambiguous in meaning
or direction. Evidence synthesis determines whether the
source supports an assertion, while ontology validation
determines whether it can be represented consistently. A
supported assertion may lack a valid ontology
representation, and an ontology-valid label does not prove
evidence support.

\noindent\textbf{Why endpoint resolution must precede
dependency fusion (RQ2).}
A supported interaction cannot become a reliable graph edge
until both endpoints are linked to the correct canonical
entities. The conservative alignment policy keeps uncertain
assertions available for review instead of using fuzzy
endpoint assignment. Although this reduces automatic
recovery, it limits false edges that could distort downstream
analysis.

\noindent\textbf{Why fusion should not require majority
agreement (RQ3).}
Most canonical dependencies were supported by only one
model. Requiring majority agreement would therefore discard
nearly all recovered dependencies. The framework instead
retains evidence-supported single-model assertions and uses
cross-model agreement as additional review metadata.
However, this does not establish that single-model
assertions are correct or that fusion outperforms the
strongest individual model.

\noindent\textbf{Why human validation remains necessary
(RQ4).}
The reference comparison shows that the automated IKB is a
review artifact rather than a validated operational graph.
Its value lies in connecting each canonical assertion to its
evidence, model contributions, ontology decisions, and
conflicts so that analysts can inspect and correct it. The
review IDG contains assertions under inspection, whereas the
validated IDG contains only human-retained assertions that
satisfy the endpoint and export rules.

\noindent\textbf{Limitations.}
The evaluation covers three open-weight models, nine project
documents, and three broad infrastructure sectors, with one
registered evidence artifact per project. The observed
behavior may not generalize to other models, sectors, or
larger evidence collections. The reference IKB was
constructed through manual document review and may reflect
annotator judgment in ambiguous cases. Independent
multi-annotator agreement has not yet been measured, and
exact alias-aware matching may miss valid semantic
equivalences that are absent from the declared aliases. Human validation is also incomplete, so the study does not
report a full comparison before and after review. In
addition, we do not compare against a direct single-pass
LLM-to-graph baseline or evaluate each model independently
against the reference. A systematic analysis of unmatched
assertions is also needed to distinguish extraction and
resolution errors from valid assertions omitted from the
conservative reference.

\noindent\textbf{Practical implications.}
Within these limits, the framework provides a feasible way
to organize heterogeneous infrastructure evidence into a
traceable and human-reviewable IKB. It does not demonstrate
fully autonomous IDG generation. Instead, it supports
analysts by making uncertain knowledge visible,
reviewable, and correctable before it enters a validated
graph. Local execution also allows sensitive OT
configuration, topology, and operational evidence to remain
within the assessed environment, which is important when
external transmission of infrastructure information is
restricted. Further improvements in entity resolution,
endpoint alignment, and relationship identification are
needed before the automated output can support operational
graph analysis with limited human review.
\section{Conclusion}
\label{sec:conclusion}

We presented an evidence-first multi-LLM framework for
constructing provenance-preserving IKBs and IDGs from
heterogeneous critical-infrastructure evidence. The
framework separates extraction, ontology grounding,
synthesis, resolution, alignment, fusion, and human
validation, and projects only human-retained knowledge into
the final IDG. Evaluation across nine projects showed that entity recovery
is easier than directed dependency recovery, with endpoint
resolution and relationship identification remaining the
main constraints. Limited cross-model support for
dependencies also supports using model agreement as review
metadata rather than a fusion requirement. The automated
IKB should therefore be treated as a traceable review
artifact, not a validated operational graph. Local execution
keeps sensitive OT evidence within the assessed environment.
Future work will improve resolution, compare individual
models and simpler baselines, evaluate completed human
review, and assess validated IDGs in downstream dependency
analysis.

\balance
\bibliographystyle{IEEEtran}
\bibliography{ref}

\newpage
\appendices
\section{Project-Level Evaluation Results}
\label{app:project_results}

Tables~\ref{tab:app_entity_results} and
\ref{tab:app_dependency_results} report project-level
pre-human agreement with the human-annotated reference IKB.
The overall rows report micro-averaged results. Aggregate
directed-endpoint agreement is reported in
Table~\ref{tab:primary_accuracy}.

\begin{table}[!h]
\centering
\caption{Project-level entity-identity agreement with the
human-annotated reference IKB.}
\label{tab:app_entity_results}
\footnotesize
\setlength{\tabcolsep}{2pt}
\renewcommand{\arraystretch}{1.05}
\begin{tabular}{@{}lrrrrrr@{}}
\toprule
\textbf{Project} &
\textbf{TP} &
\textbf{FP} &
\textbf{FN} &
\textbf{P} &
\textbf{R} &
\(\boldsymbol{F_1}\) \\
\midrule
AusNet Gas
  & 15 & 178 & 4
  & 7.77 & 78.95 & 14.15 \\

Boys Water
  & 20 & 33 & 11
  & 37.74 & 64.52 & 47.62 \\

Iqaluit
  & 18 & 902 & 12
  & 1.96 & 60.00 & 3.79 \\

MANK2
  & 37 & 91 & 11
  & 28.91 & 77.08 & 42.05 \\

Parris Island
  & 8 & 120 & 8
  & 6.25 & 50.00 & 11.11 \\

Pyramid NG
  & 28 & 130 & 12
  & 17.72 & 70.00 & 28.28 \\

Snoqualmie
  & 12 & 617 & 10
  & 1.91 & 54.55 & 3.69 \\

TigerSkid
  & 45 & 311 & 9
  & 12.64 & 83.33 & 21.95 \\

Washington IL
  & 14 & 113 & 10
  & 11.02 & 58.33 & 18.54 \\
\midrule
\textbf{Micro}
  & \textbf{197}
  & \textbf{2495}
  & \textbf{87}
  & \textbf{7.32}
  & \textbf{69.37}
  & \textbf{13.24} \\
\bottomrule
\end{tabular}
\end{table}

\begin{table}[!h]
\centering
\caption{Project-level agreement for complete dependency
relationships. A match requires the correct canonical
source, relationship, and target in the correct direction.}
\label{tab:app_dependency_results}
\footnotesize
\setlength{\tabcolsep}{2pt}
\renewcommand{\arraystretch}{1.05}
\begin{tabular}{@{}lrrrrrr@{}}
\toprule
\textbf{Project} &
\textbf{TP} &
\textbf{FP} &
\textbf{FN} &
\textbf{P (\%)} &
\textbf{R (\%)} &
\(\boldsymbol{F_1}\) \textbf{(\%)} \\
\midrule
AusNet Gas
  & 0 & 20 & 7
  & 0.00 & 0.00 & 0.00 \\

Boys Water
  & 0 & 2 & 10
  & 0.00 & 0.00 & 0.00 \\

Iqaluit
  & 0 & 24 & 10
  & 0.00 & 0.00 & 0.00 \\

MANK2
  & 1 & 12 & 13
  & 7.69 & 7.14 & 7.41 \\

Parris Island
  & 0 & 14 & 8
  & 0.00 & 0.00 & 0.00 \\

Pyramid NG
  & 0 & 32 & 20
  & 0.00 & 0.00 & 0.00 \\

Snoqualmie
  & 0 & 55 & 14
  & 0.00 & 0.00 & 0.00 \\

TigerSkid
  & 0 & 51 & 12
  & 0.00 & 0.00 & 0.00 \\

Washington IL
  & 0 & 7 & 7
  & 0.00 & 0.00 & 0.00 \\
\midrule
\textbf{Micro}
  & \textbf{1}
  & \textbf{217}
  & \textbf{101}
  & \textbf{0.46}
  & \textbf{0.98}
  & \textbf{0.63} \\
\bottomrule
\end{tabular}
\end{table}

\section*{LLM Usage Statement}

\noindent\textbf{Editorial use.}
LLMs were used for editorial purposes in this manuscript, and all outputs were inspected by the authors to ensure accuracy and originality.

\end{document}